\documentclass[twocolumn]{aastex631}

\newcommand{\Msun}{M$_\odot$} 
\newcommand{\Rsun}{R$_\odot$} 
\newcommand{\Lsun}{L$_\odot$} 

\usepackage{amsmath}
\usepackage{comment}

\newif\ifref
\reftrue
\reffalse
\definecolor{darkred}{rgb}{0.75, 0, 0}
\newcommand{\mb}[1]{\ifref\textcolor{darkred}{#1}\else #1\fi}

\accepted{October 31, 2024}

\begin{document}

\title{Formation of Sub-Chandrasekhar Mass Black Holes and Red Stragglers via Hawking Stars in Ultra-Faint Dwarf Galaxies}

\author{Andrew D. Santarelli}
\affiliation{Department of Physics, Illinois State University, IL, USA}

\author{Matthew E. Caplan}
\affiliation{Department of Physics, Illinois State University, IL, USA}

\author{Earl P. Bellinger}
\affiliation{Department of Astronomy, Yale University, CT, USA}

\begin{abstract}

Presently, primordial black holes (PBHs) in the asteroid-mass window from $10^{-16}$~\Msun\ to $10^{-10}$~\Msun\ are a popular dark matter candidate. If they exist, some stars would capture them upon formation, and they would slowly accrete the star over gigayears.
Such Hawking stars -- stars with a central PBH -- provide a novel channel for the formation of both sub-Chandrasekhar mass black holes as well as red straggler stars. 
Here we report on stellar evolution models that extend our previous work to Hawking stars with masses between 0.5 and 1.4~\Msun. 
We explore three accretion schemes, and find that a wide range of PBHs in the asteroid-mass window can robustly accrete stars as small as 1~M$_{\odot}$ within the age of the Universe. 
This mechanism of producing sub-solar mass black holes is highly dependent on the assumed accretion physics and stellar metallicity. 
Lower-metallicity stars are generally accreted more rapidly, suggesting that it may be more likely for sub-Chandrasekhar mass Hawking stars formed in the early universe, such as those in ultra-faint dwarf (UFD) galaxies, to transmute their star into a sub-Chandrasekhar mass black hole within a Hubble time. 
We present a stellar population synthesis of a Draco II-like UFD galaxy containing Hawking stars and show that the number of red stragglers they produce can qualitatively match the observed population \mb{for black hole seed masses around 10$^{-11}$~\Msun\ and under the assumption that they accrete with high radiative efficiency}. 
\end{abstract}


\section{Introduction} \label{sec:intro}
 
The only known production mechanisms for black holes in the present universe are neutron star mergers and supernovae, both producing black holes with masses greater than the Chandrasekhar limit, roughly 1.4~\Msun. One expects stars with degenerate cores exceeding this to explode in core collapse supernovae, either producing neutron stars or black hole. Thus, any black hole with a mass less than the observed upper-limit for neutron stars very likely has a non-stellar origin, while black holes less than a Chandrasekhar mass almost certainly have an exotic formation mechanism. \mb{Other variables such as rapid rotation and neutron star mergers may also be an important factor, though these may more likely result in subsolar-mass neutron stars rather than black holes \citep{metzger2024}}. The LIGO-Virgo-Kagra gravitational wave collaboration is sensitive to merging black holes of these masses, and searches for these objects are underway \citep{lvk2023search}.

Among the most popular alternate formation channels for black holes are density fluctuations in the early universe that produce primordial black holes (PBHs) \citep{Hawkins2011,Green2021,Carr2022,Carr2024}. Such black holes do not have their masses constrained by stellar physics so a \emph{bona fide} detection of a sub-Chandrasekhar mass black hole would be strong, if not definitive, evidence for PBHs. PBHs have become especially popular in recent years as a dark matter candidate, though large mass ranges are now tightly constrained by observations \citep{Carr2022}. Presently, the `asteroid' mass window between $10^{-16}$~\Msun\ and $10^{-10}$~\Msun\ remains largely unconstrained and could still constitute all of the dark matter. 

If PBHs in this mass range exist and are a large fraction of the dark matter, then some would inevitably become captured by stars.
\cite{Hawking1971} provided the first order-of-magnitude estimate for the stellar PBH capture rate, suggesting that a PBH of $\sim 10^{17}$~g could be captured by a Sun-like star. \cite{Capela2013} considered capture rates in globular clusters and the survival of white dwarfs and neutron stars as a potential constraint on PBHs. \cite{Oncins2022} also considered capture in early, low-metallicity dwarf galaxies, finding the rate is largely insensitive to the PBH mass in the asteroid-mass window, and that many early stars could have been transmuted to sub-solar mass black holes.
Recently, \cite{esser2023constraints} and \cite{esser2023impact} computed the capture rate of star forming clouds from detailed numerical simulations of adiabatic cloud contraction. They show that the capture probability scales strongly with $\rho_{\mathrm DM}$ the local dark matter density and $\sigma^{3}$ the velocity dispersion cubed, resulting in capture rates of order one in some ultra-faint dwarf galaxies (UFDs), with capture probabilities roughly of order $10^{-7}$ in Milky Way-like galaxies \citep{Caplan2023}. Therefore, Hawking stars (and by extension sub-solar mass black holes produced by them) may be common in UFDs but rare in the disk of the Galaxy. 

If a low mass star captures a PBH it could accrete the star within the age of the Universe, providing a formation pathway for sub-Chandrasekhar mass black holes. 
These stars would be partially powered by black hole accretion in their cores as they are accreted over gigayears, and have been called `Hawking stars' \citep{Bellinger2023,Caplan2023}. Assuming the core black hole accretes the entire star without significant mass loss, the Hawking star transmutes an asteroid-mass PBH into a stellar mass black hole. Even in the absence of PBHs, it is also worth noting that some particle dark matter candidates could also produce Hawking stars if a sufficiently large dark matter halo becomes bound to a star and relaxes to form a compact core \citep[e.g.][]{Ge2024}.

\cite{Bellinger2023} and \cite{Caplan2023} developed new capabilities for the 1D stellar evolution code \textsc{Mesa} \citep{Paxton2011, Jermyn_2023} for simulating Hawking stars and reported numerical simulations of solar-mass Hawking stars. These models show that Hawking stars can be surprisingly long-lived due to the low accretion rates onto such small black holes, with some stars living out their entire main sequence life before the black hole comes to dominate the evolution of the star. When the black hole luminosity exceeds the solar luminosity, roughly around black hole masses of $10^{-6}$~\Msun, the star expands, becomes fully convective, and fusion quenches. The surface properties of the post-main sequence Hawking star are consistent with a `red straggler,' at roughly 10~\Lsun and 4~\Rsun for several gigayears before the star is fully accreted \citep{Bellinger2023}. Depending on the accretion model, Hawking stars provide a formation pathway for red stragglers seen in CMDs of UFD galaxies such as Draco II \citep{weisz2023jwst}. We are therefore motivated to study the lifetime of Hawking stars for a range of star and PBH seed masses to determine the viability of Hawking stars as a formation channel for both sub-Chandrasekhar mass black holes and red stragglers. 

In this work, use the methods of \cite{Bellinger2023} and \cite{Caplan2023} to study other sub-Chandrasekhar stellar masses that capture PBHs upon formation and estimate the lifetimes of Hawking stars. We begin with a description of our accretion schemes in Sec.~\ref{sec:acc} and present analytic estimates of lifetimes in Sec.~\ref{sec:analytic}. We present numerical results to verify and extend these calculations in Sec.~\ref{sec:SEM} using new \textsc{Mesa} models of Hawking stars across a range of masses. In Sec.~\ref{sec:popsyn} we use these \textsc{Mesa} models for population synthesis simulations of a Draco II-like UFD galaxy containing Hawking stars and compare to recent observations.

\section{Accretion Schemes}\label{sec:acc}

\begin{figure}
    \centering
    \includegraphics[trim={0 0 0 0},clip,width=0.46\textwidth]{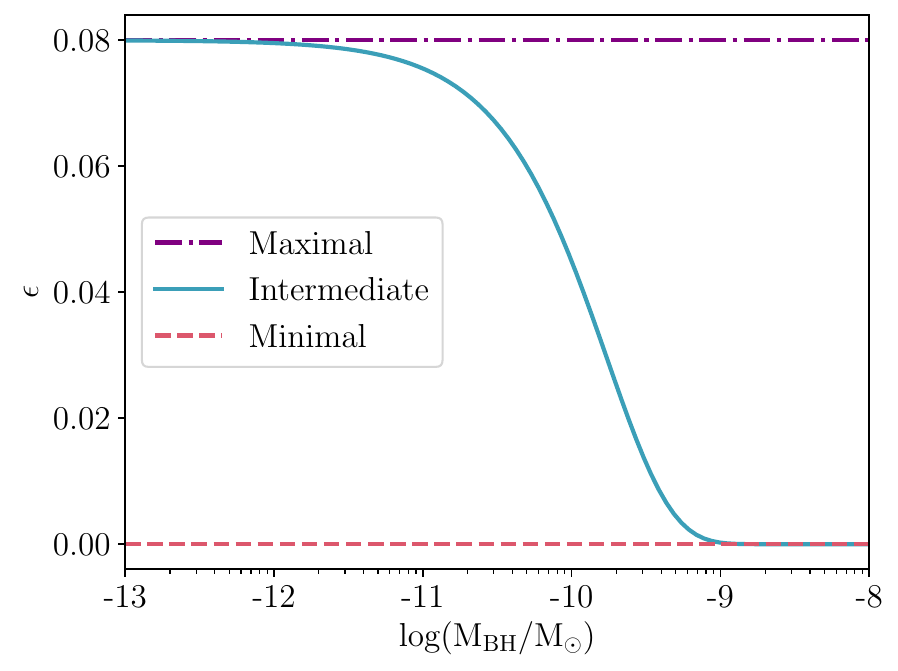}
    \caption{Maximal (purple, dotted), intermediate (blue, solid), and minimal (red, dashed) radiative efficiencies as a function of black hole mass. Maximal and minimal schemes are both constant, maximal being that found in \cite{Bellinger2023} and minimal corresponding to the canonical Bondi accretion scheme.}
    \label{fig:epsilon}
\end{figure}

We begin with a description of our accretion schemes and stellar evolution models of Hawking stars. Our \textsc{Mesa} models treat the PBH seed as a point mass at the center of the star and introduce the new boundary condition $M(R=0)=M_{\rm PBH}$, and then solve the 1D equations of stellar structure. The BH accretes from the Bondi sphere with a radius
\begin{equation}
\label{R_B}
    R_B=\frac{2GM_{\mathrm{BH}}}{c_s^2}
\end{equation}
where $c_s$ is the sound speed of the medium at $R_B$. \mb{Inside the Bondi sphere, we take $\rho \propto r^{-3/2}$ under the assumption that the viscosity from small scale magnetic fields is large enough to transport angular momentum outward while convection transports it inward, thus there is finite mass flux onto the black hole and zero net angular momentum and energy flux \citep{Bondi1952, Quataert_2000, ball2012quasistars}. This density profile then gives} a mass $(8\pi /3)\rho R_B^3$ contained within $R_B$. This is small compared to the BH and allows us to treat the Bondi sphere as a cavity that the BH accretes from. This solves many of the numerical issues at the core and prevents the density cusps associated with accretion. However, this requires us to use \emph{ad hoc} accretion rates and luminosities. Self-consistent accretion physics is beyond the scope of our stellar evolution models and likely requires detailed (possibly multidimensional) subgrid modeling. Instead, we explore the effects of several different accretion schemes that we construct analytically that explore a range of possible physics. 

The lifetime of a Hawking star is heavily dependent on accretion physics that remain poorly understood. Even with recent advances in numerical computing, exact expressions for accretion rates and luminosities -- especially for microscopic PBHs accreting from the cores of stars -- are not readily available in the literature. In fact, this has been an outstanding problem for decades \citep{1981MNRAS.194..475T}. In lieu of this, we have in our past work developed several accretion schemes. We begin with the definition of the accretion luminosity 
\begin{equation}
\label{eq:lum}
    L=\frac{\epsilon}{1-\epsilon} \, \dot M_{\mathrm{BH}} \, c^2
\end{equation}
which depends on the mass accretion rate $\dot M_{\mathrm{BH}}$ and radiative efficiency $\epsilon$. Any two variables of \mb{$\dot M_{\mathrm{BH}}$}, $L$, or $\epsilon$ set the third. 

Our first accretion scheme, developed by \cite{ball2012quasistars} and expanded on for Hawking stars in \cite{Bellinger2023}, is our `maximal' scheme where we assume ${\epsilon = 0.08}$, \mb{a typical radiative efficiency for a Schwarzschild black hole corresponding to  the energy extracted from accreting matter reaching its innermost stable orbit}. We take the BH luminosity to be the lesser of the convection-limited Bondi luminosity and the Eddington limit
\begin{align}
    \label{eq:Ledd}
    L_{\mathrm{E}} &= 4\pi \frac{c}{\kappa} GM_{\mathrm{BH}},\\
    \label{eq:Lbondi}
    L_{\mathrm{B}} &= 16\pi \eta \frac{\rho}{c_s \Gamma_1} (GM_{\mathrm{BH}})^2
\end{align}
where ${\eta=0.1}$ is the convective efficiency, ${\Gamma_1=5/3}$ is the first adiabatic exponent, and $\kappa$ is the opacity which changes with stellar mass and metallicity. \mb{Almost all of the flux in this case is carried away via convection, the velocity of which can be no more than the sound speed without dissipating its energy to shocks. The convective flux is then, at most, order $pc_s$ which results in a reduction to the canonical Bondi luminosity by a factor of $4c_s^2/\Gamma_1\lambda_c$  \citep{ball2012quasistars}, where $\lambda_c$ is a factor dependent on $\Gamma_1$ \citep{Bondi1952}. This reduction results in Eq. \ref{eq:Lbondi}.} The accretion rate is then found from Eq.~\ref{eq:lum}. 
\mb{When the accretion luminosity exceeds the core fusion luminosity, the area outside the Bondi sphere is heated via the black hole radiation, which causes the star to expand and the accretion rate to decrease. It is at this point that the Hawking star evolves into a long-lived subgiant and then red straggler,} which was studied in detail in our past work. 

For our second scheme developed in \cite{Caplan2023}, the `intermediate' scheme, we use an adaptive radiative efficiency to approximate the effects of photon trapping. This scheme assumes that photons only escape \mb{when the free-fall time exceeds the photon diffusion time at the Bondi radius.} The radiative efficiency therefore becomes a function of the BH mass 
\begin{equation}
\label{eq:eps}
    \epsilon=\epsilon_0 \, \mathrm{exp} \left [  -\frac{GM_{\mathrm{BH}} \rho \kappa}{\pi c c_s} \right ]
\end{equation}
such that the BH accretion luminosity goes to zero when $M_{\rm BH} \gtrsim \pi c c_s/G \rho \kappa $. For values typical of the solar core, this transition occurs around $10^{-10}$~\Msun\ and can be seen in Fig.~\ref{fig:epsilon}. We take $\epsilon_0 = 0.08$ so that that black hole radiates efficiently at low masses, but then transitions to trapped accretion with $\epsilon \approx 0$ over roughly an order of magnitude in mass growth, shown in Fig.~\ref{fig:epsilon}. As before, the black hole luminosity is taken to be the minimum of the convective limit and the Eddington limit when $\epsilon \approx 0.08$, but as $\epsilon \rightarrow 0$ the mass accretion rate would diverge to infinity, so we place an upper limit on the mass accretion rate of the canonical Bondi rate,
\begin{equation}
\label{eq:canon}
    \dot M_{\mathrm{BH}} = \frac{\pi \rho G^2 M_{\mathrm{BH}}^2}{c_s^3}.
\end{equation}
\mb{Once accreting at this upper limit (for fixed BH mass) and the radiative efficiency is effectively zero, the luminosity rapidly goes to zero and the star is rapidly consumed without the red straggler phase seen in the previous scheme.}
Due to numerical roundoff errors, we set a floor at $\epsilon=1.1\times 10^{-20}$ in our \textsc{Mesa} models. See Fig.~3 of \cite{Caplan2023} for more details. 

Our final accretion scheme is the `minimal' accretion scheme in which photons are effectively always trapped and the accretion rate is the canonical Bondi rate of Eq.~\ref{eq:canon} with $\epsilon =0$ for all BH masses. This is a common choice by many authors, \mb{though perhaps not the most realistic compared to the other schemes and serves as an upper bound on the accretion rate for our purposes}. If one takes the mass density in the Bondi sphere to be $\rho \propto r^{-3/2}$ then the photon mean free path $\ell = (\kappa \rho(r))^{-1}$ becomes a strong function of $r$. The diffusion timescale becomes long at small $r$ while the freefall time is short, resulting in an effective trapping radius close to the BH. For a $5\times10^{-14}$~\Msun\ BH in the solar core this trapping radius is of order 200 Schwarzschild radii. As most of the accretion luminosity is released within a few Schwarzschild radii of the BH, this calculation suggests that almost all luminosity is trapped so that $\epsilon$ is small \citep{EsserPC}.   

\mb{We emphasize that these models are all simplistic and are chosen to bracket the possible physics rather than to present a model derived from first principles microphysics. Further progress requires detailed general relativistic magnetohydrodynamics (GRMHD) simulations to determine the radiative efficiency. For example, our models treat the inflow within the Bondi sphere as a black box, and suppose thermal equilibrium between photons and infalling matter. As the plasma infalls, the temperature increases as $r^{-1}$ while the photon temperature increases as $r^{-1/2}$, requiring the production of photons by some mechanism such as Compton scattering or bremsstrahlung. We do not account for this, so there may be further limits on the luminosity due to the production timescale of these photons.}

We remain agnostic to which of these accretion schemes is the most physically realistic. None of these accretion schemes are particularly implausible for 1D accretion, though important physics has been neglected including angular momentum, magnetic fields, and multidimensional flows including disks and jets (see Sec.~4 of \citealt{Bellinger2023}). Given the wide range of stellar masses, rotation rates, and metallicities, it seems possible that some Hawking stars may be well described by the maximal scheme while others are well described by an intermediate or minimal scheme. This question is ultimately beyond the scope of this work, but our extreme sensitivity to accretion schemes is an important caveat and we hope to motivate hydrodynamic calculations of the accretion rates and luminosities of microscopic BHs.

\section{Lifetimes} \label{sec:lifetimes}
\subsection{Analytic Lifetimes}\label{sec:analytic}

\begin{figure*}
    \centering
    \includegraphics[trim={0 0 5cm 0},clip,height=0.32\textwidth]{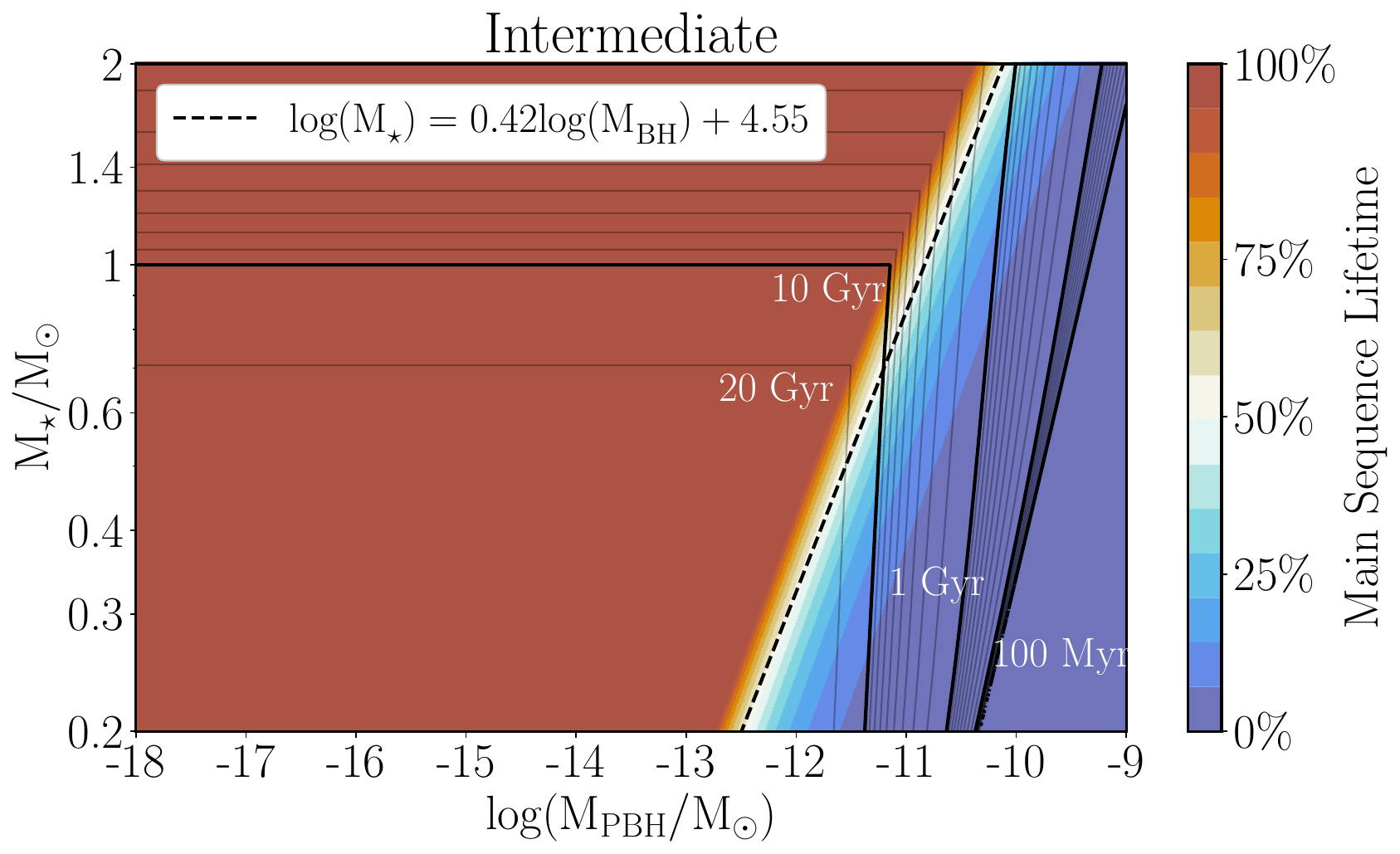}
    \includegraphics[trim={0 0 0 0},clip,height=0.32\textwidth]{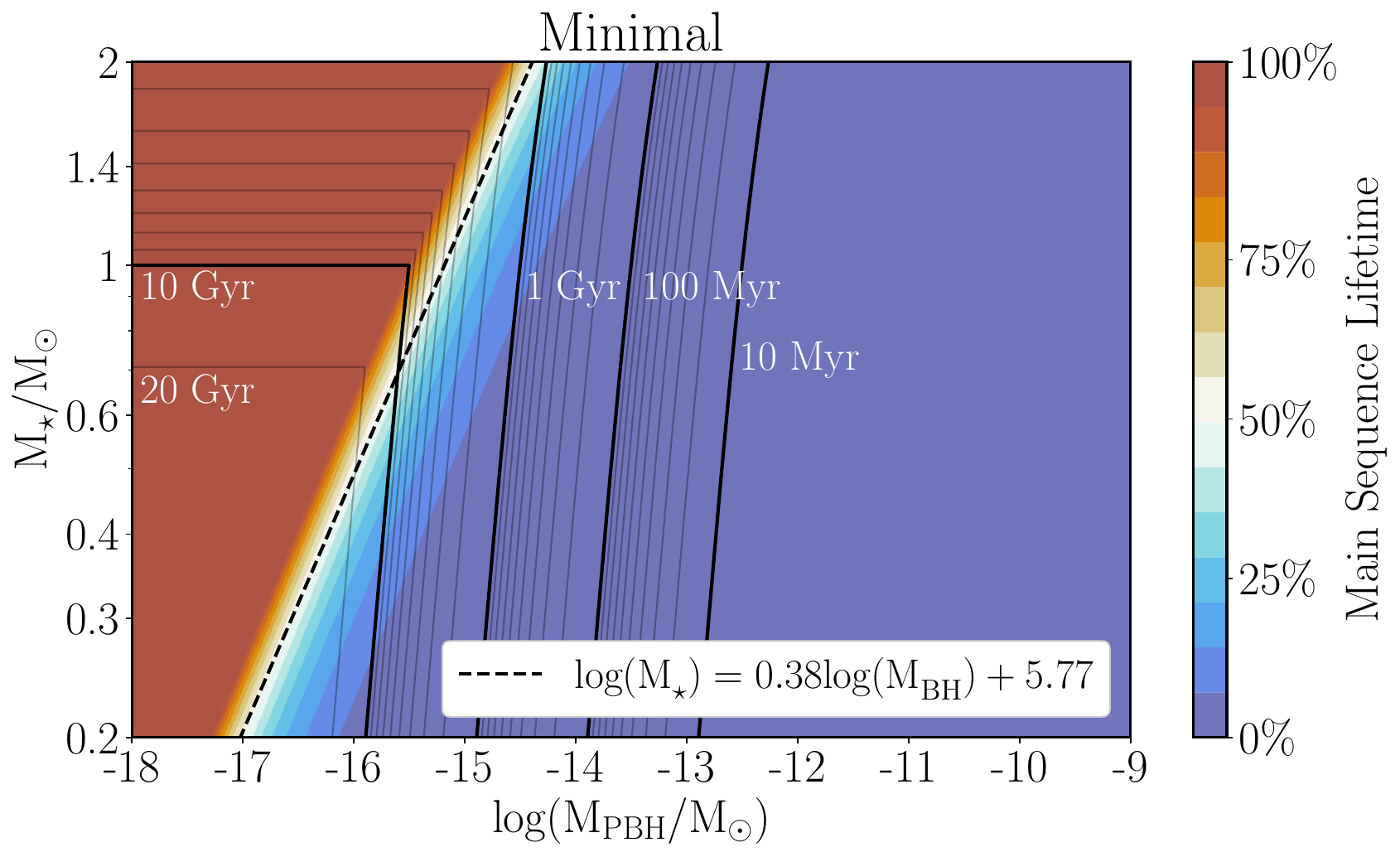}
    \caption{Hawking star lifetimes as a percentage of the typical main sequence lifetime, all as a function of both the stellar and PBH seed masses. The intermediate radiative efficiency scheme is used for the left plot, and the minimal for the right. Solid black lines indicate actual age in log scale, and the dotted black line is fit to the 50\% main sequence lifetime curve. The maximal scheme can be found in Fig.~4 of \cite{Bellinger2023}.}
    \label{fig:bondipepsi}
\end{figure*}

We begin by making analytic estimates of Hawking star lifetimes as a percentage of the main sequence lifetime of a star without a central PBH. We use estimates the for zero age main sequence (ZAMS) values for $\kappa$, $\rho$, and $c_s$ from the appendix of \cite{Bellinger2023},
\begin{align}
    c_s &= 1.6 \times 10^7\, (3.5 + \log_{10}M_\ast) \label{eq:cs}
    \, ,\\
    \kappa &= 0.3 + 1.2\times M_\ast^{-1.3}
    \, \label{eq:kappa} ,\\
    \log_{10} \rho &= \dfrac{-1.9}{1 + \exp\{- 2.7 (\log_{10} (M_\ast) - 1) \}} + 2
    \, ,
\end{align}
where $M_\ast$ is in \Msun\ units and the rest are cgs. Below we compute the lifetimes for the intermediate and minimal schemes; the estimates and detailed analysis for the maximal radiative efficiency scheme can be found in \cite{Bellinger2023}. 

In the case of the intermediate scheme, we break our calculation into two parts. We first estimate the time until the Bondi-Eddington transition, where we assume a constant radiative efficiency $\epsilon=0.08$ as it remains largely unchanged until said transition. For this we use

\begin{equation}\label{eq:lt-int}
\begin{split}
    t_{\mathrm{Bondi}}&=S_\mathrm{Bondi}\int_{M_\mathrm{BH,0}}^{M_\mathrm{BET}} \frac{1}{M_\mathrm{BH}^2} dM_\mathrm{BH} \\
    &=-S_{\mathrm{Bondi}} \left ( \frac{1}{M_{\mathrm{BET}}} - \frac{1}{M_{\mathrm{BH,0}}} \right ),
\end{split}
\end{equation}

\begin{equation}
    M_\mathrm{BET}=\frac{cc_s \Gamma_1}{4G \rho \kappa \eta},
\end{equation}

\begin{equation}\label{eq:Sbondi}
    S_{\mathrm{Bondi}} = \frac{\epsilon}{(1-\epsilon)\eta} \frac{c^2 \Gamma_1 c_s}{16\pi G^2 \rho},
\end{equation}

\noindent where $M_\mathrm{BH,0}$ is the PBH seed mass, $M_\mathrm{BET}$ is the mass at which the accretion transitions from \mb{convection-limited} Bondi to Eddington, and $S_{\mathrm{Bondi}}$ is the mass sensitivity for convection-limited Bondi accretion. We then ignore the proceeding Eddington-limited accretion time as it lasts order $10^3$ years which is within error \citep{Caplan2023}, and \mb{transition to} the canonical Bondi scheme until the star has been completely consumed. This is a simple change to the boundaries of Eq.~\ref{eq:lt-int}, integrating from $M_\mathrm{BET}$ to $M_\ast$ instead. We must also change the mass sensitivity $S_\mathrm{Bondi}$ to that of the canonical Bondi regime, given by
\begin{equation}
    S_\mathrm{Bondi,0}=\frac{c_s^3}{\pi G^2 \rho}.
\end{equation}
In the case of our ``minimal'' scheme, \mb{we simply use the canonical Bondi lifetime calculation but for the entire growth of the black hole, i.e.} integrated from $M_\mathrm{BH,0}$ to $M_\ast$.

We use these methods to estimate the complete accretion timescales for a range of stellar and PBH seed masses with values 0.2 to 2 and $10^{-18}$ to $10^{-9}$ solar masses respectively. We see on the left plot of Fig.~\ref{fig:bondipepsi} that the main sequence lifetimes of subsolar mass stars begin to get cut short for PBH seed masses in the range of about $10^{-13}$ to $10^{-11}$ solar masses, firmly within the currently unconstrained asteroid mass window. However, only those with seed masses greater than about $3 \times 10^{-12}$ solar masses would accrete fully during their main sequence lifetimes within the age of the Universe. Notably, stars above about 0.9 solar masses can be transmuted within the age of the Universe due to core contraction upon entering the post main sequence, leading to rapid accretion. 

In contrast, the right plot of Fig.~\ref{fig:bondipepsi} shows the lifetimes under our ``minimal'' regime where only canonical Bondi accretion is used. The importance of the convective limit throttling the accretion (in the maximal and intermediate schemes) is immediately apparent, as now PBH seed masses as low as the evaporation constraints ($\sim 10^{-16}$~\Msun) can completely consume stars as small as 0.2~\Msun\ within the age of the Universe. 

The expected post-main sequence evolution of the Hawking star can also be read off of these plots. Shown in red are stars with sufficiently small PBHs that finish their canonical main sequence and begin a normal red giant phase before being rapidly consumed and converted to a stellar mass black hole during core contraction. Those Hawking stars that grow to become the dominant source of luminosity before the end of the main sequence and evolve into red stragglers that take Gyrs to accrete are shown in orange, yellow, white, and blue \citep{Bellinger2023}. 

This also suggests that metallicity effects may be important, as small changes in metallicity can drastically impact the core opacity, sound speed, and density and thus the exact accretion rates at low stellar masses. Similarly, the dominance of CNO burning above a solar mass can drastically change the main sequence lifetime of more massive stars, and must be studied in detail with stellar evolution models. 
\subsection{Stellar Evolution Models}\label{sec:SEM}

\begin{figure*}
    \centering
    \includegraphics[width=0.9\textwidth]{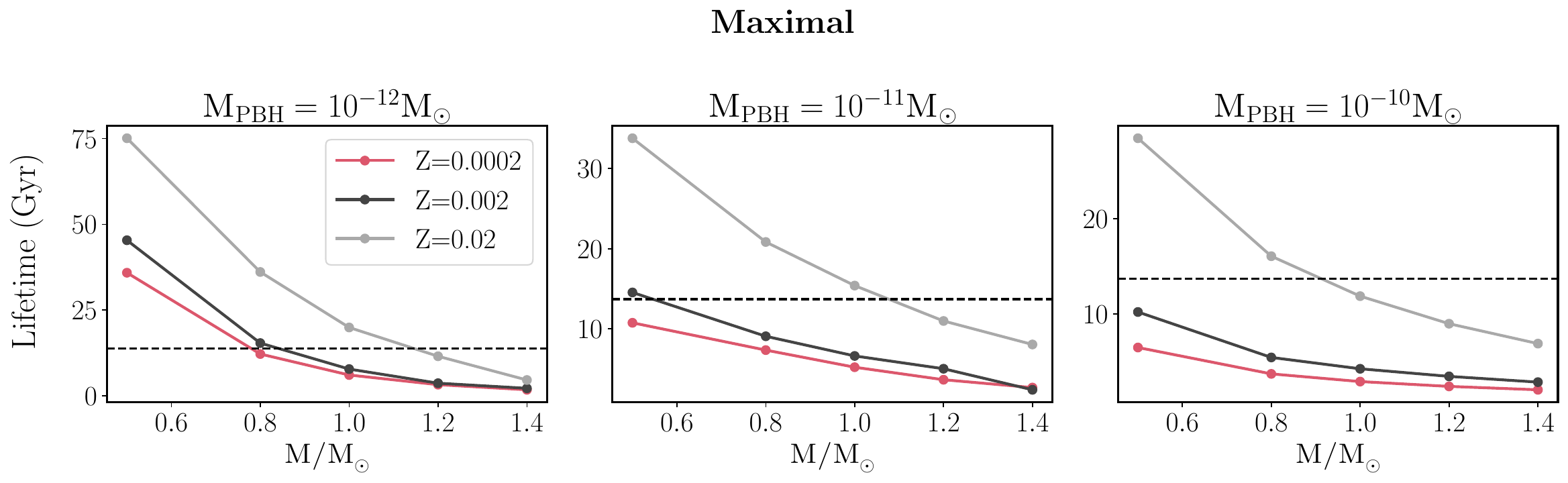}
     \vspace{2.5 mm} \hrule \vspace{2.5 mm} 
    \includegraphics[width=0.9\textwidth]{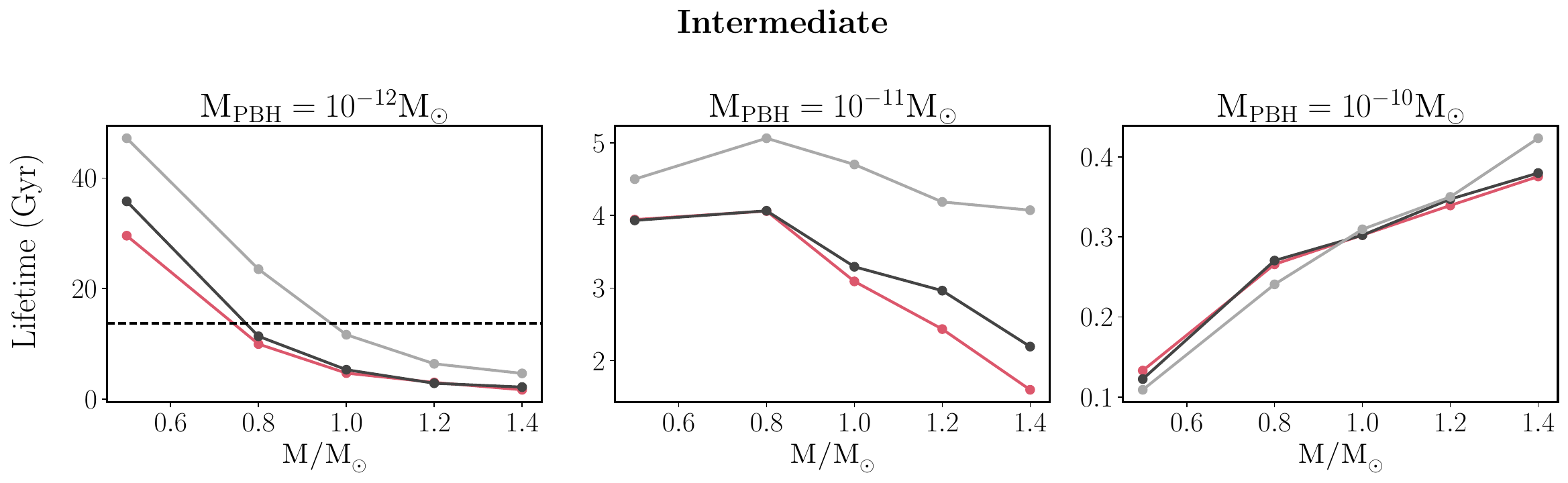}
     \vspace{2.5 mm} \hrule \vspace{2.5 mm} 
    \includegraphics[width=0.9\textwidth]{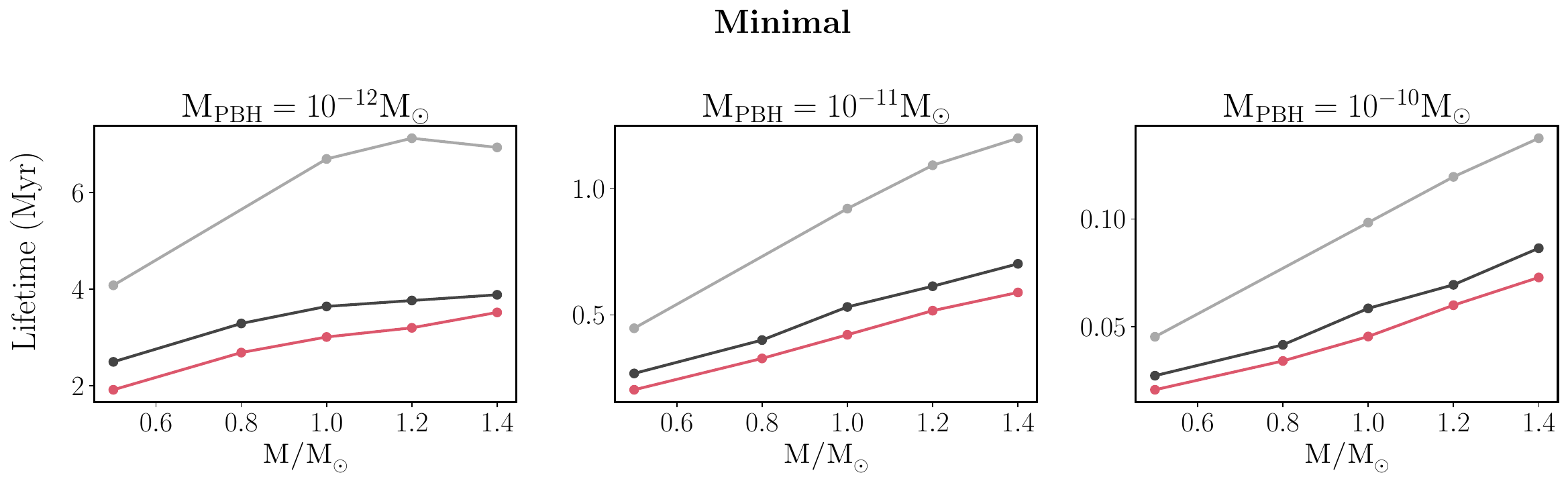}
    \caption{Lifetime of Hawking stars in our \textsc{Mesa} models with each accretion scheme as a function of stellar mass for PBH seed masses $10^{-12}$\Msun\ (left), $10^{-11}$\Msun\ (middle), and $10^{-10}$\Msun\ (right) for 1\% (red), 10\% (black), and 100\% (gray) solar metallicity. The dashed black line indicates a Hubble time 13.7~Gyr where appropriate and is excluded on plots with lifetimes well below it. Note that the lifetimes for the minimal regime are in Myr as opposed to Gyr for the other two.}
    \label{fig:num-lt}
\end{figure*}

Our generalized analytic calculations serve as motivation to numerically model these sub-Chandrasekhar mass Hawking stars to calculate lifetimes and post main sequence evolution, and in doing so we can also compare lifetimes across metallicities. For these models, we implement the three accretion schemes outlined in Sec.~\ref{sec:acc} in \textsc{Mesa}. 

We run \textsc{Mesa} models for a range of stellar masses, PBH seed masses, and metallicities - the lifetimes of these Hawking stars are shown in Fig.~\ref{fig:num-lt}. The Hawking star lifetimes at solar metallicity in Fig.~\ref{fig:num-lt} agree with our analytic calculations within a factor of two, with the difference dominated by metallicity. We choose stellar masses of 0.5~\Msun\ to 1.4~\Msun\ because in this work we are most interested in the production of subsolar and sub-Chandrasekhar mass black holes. We consider PBH seed masses of $10^{-12}$~\Msun\, $10^{-11}$~\Msun\, and $10^{-10}$~\Msun\ because our analytic calculations with the intermediate model above motivates a closer study of these masses; in addition, these are near the upper end of the asteroid-mass window. 
We explore three stellar metallicities, 1\%, 10\%, and 100\% of solar metallicity, in order to understand how Hawking star lifespans may evolve with cosmic time, and especially because the PBH capture rate may have been higher in the low mass, low metallicity halos of the earlier Universe \citep[e.g.][]{Lokas2001,Oncins2022}. 

In the intermediate and minimal models, we define the lifetime as the time at which the black hole mass has reached 0.1\% of the stellar mass, as this is within hours of complete accretion and we do not think the stellar evolution models are reliable beyond that point due to numerical errors and limitations of our accretion schemes. In the event the BH never reaches this threshold, we use the final age in the simulation as the consumption time as the core contraction in the post main sequence leads to several orders of magnitude growth in the accretion rates due to the greater core conductivities, densities, and sound speeds \citep{Caplan2023, Bellinger2023}. The exception to this threshold is in our maximal scheme as the Hawking star can continue to live through a prolonged sub-subgiant phase after it is black hole dominated \citep{Bellinger2023}. With that and post main sequence core contraction taken into account, we can use the final age as our transmutation time for all of these models since they run to completion and do not experience the aforementioned errors.

For most cases considered in Fig.~\ref{fig:num-lt}, we find that almost all low metallicity Hawking stars with a PBH seed mass of at least $10^{-12}$~\Msun\ are consumed by their black holes within the current age of the Universe, the only exceptions being those with stellar masses below 0.8~\Msun\ in the maximal regime. A large stellar mass dependence is seen in all models, though the trend with mass is model dependent (discussed below). Perhaps more interestingly, there is a strong metallicity dependence throughout as well; a factor of two decrease in the lifetime between solar metallicity and 1\% solar metallicity is typical. This suggests that early universe Hawking stars would be more readily transmuted into black holes within a Hubble time, relative to more recently formed stars. 

We now discuss the trends in lifetime for each accretion scheme individually. We see a monotonic decay in lifetime with stellar mass for the maximal scheme. These models undergo the transition from convectively limited Bondi to Eddington limited accretion, where they remain through the rest of their lifetimes. During Eddington accretion $\dot{M}_{\rm Edd} \propto \kappa^{-1}$ with an e-folding growth timescale $\tau_{\rm Edd} \propto \kappa$ \citep{Caplan2023}. Core opacity shows a strong inverse dependence on stellar mass in Eq.~\ref{eq:kappa} which produces the dependence seen in the maximal model. As Eq.~\ref{eq:kappa} does not have any metallicity dependence, we include plots of the core properties of these models at ZAMS in the Appendix, making the scaling of lifetime on metallicity and stellar mass more clear. 

In the minimal scheme, lifetimes are significantly shorter due to the rapid growth that occurs with Bondi accretion. As was seen in the analytic estimates in Sec.~\ref{sec:analytic}, all PBH seed masses in the asteroid mass window result in transmutation to a BH within the current age of the Universe. For higher seed masses in the minimal scheme, full transmutation occurs before star formation is complete. The minimal scheme shows an opposite trend from the maximal scheme, where the lifetime now grows with mass. This can again be understood from analytics. The Bondi accretion rate is $\dot{M}_{\rm B} \propto c_s^{-3}$, and sound speed scales like $c_s \propto \log_{10}M_\ast$ in Eq.~\ref{eq:cs}, resulting in a characteristic growth timescale $\tau_{\rm B} \propto 3 \log_{10}M_\ast$. The 1.5~\Msun\ Hawking stars have a lifetime roughly 1.5 ($\approx 3 \log_{10}3$) times longer than 0.5~\Msun\ stars, consistent with this interpretation. To better understand the dependence on metallicity we include the core sound speed and density at ZAMS in the appendix, as the canonical Bondi accretion rate also scales with core density. While the core density does not vary much with stellar mass (at constant metallicity), the core densities are lower with higher metallicity, and so we find that larger metallicities can result in significant extensions of the lifespan. 

In the intermediate scheme we observe the effects of photon trapping via the change in stellar mass dependence with PBH seed mass. With lower seed masses (left), the lifetime is dominated by the early convectively limited phase before the onset of luminosity trapping and the star evolves like the maximal models for the majority of their lives. For a seed mass of $10^{-11}$~\Msun\ (middle), the stars spends comparable fractions of its life in a convectively limited regime and a canonical Bondi regime, roughly flattening the trend of lifetime with stellar mass. This mass dependence is due to the accretion's transition from a $c_s^{-1}$ to a $c_s^{-3}$ dependence, making it much more sensitive to increased stellar mass. For higher seed masses in the intermediate scheme (right), the lifetime now grows with stellar mass and metallicity dependence drops out nearly entirely, with a radiative efficiency like the minimal models for the entire life of the Hawking star. 

Even with large uncertainties in the accretion physics, we find that even Hawking stars with small seed PBHs can robustly accrete stars above a solar mass within the age of the Universe due to the canonical main sequence lifetime of these stars being less than a Hubble time. The lifetimes of lower mass Hawking stars, especially at subsolar masses, show incredibly strong sensitivity to accretion physics, PBH seed mass, and metallicity. When the accretion is luminous and throttled by the convective limit the lifetime of subsolar mass stars easily exceeds a Hubble time and so one would not expect any subsolar mass black holes to be produced by this mechanism. However, in the maximal scheme PBHs with seed masses at the upper limit of the asteroid-mass window ($10^{-10}$~\Msun) can still accrete low-mass, low-metallicity stars, so there is a pathway for Hawking stars to transmute PBHs to subsolar masses here, albeit with fine-tuning. In contrast, in our minimally luminous accretion scheme with only canonical Bondi accretion, we find that virtually any stellar mass can be accreted by any PBH mass in less than a Hubble time. \mb{This is scheme is likely unrealistic, however, and serves as an upper bound on the accretion rate for our purposes.}

\section{Population Synthesis}\label{sec:popsyn}

Ultra faint dwarf (UFD) galaxies are dark matter rich, low velocity dispersion satellite galaxies formed in the early universe, making them suitable sites for the production of Hawking stars. Using JWST data, \citet{weisz2023jwst} presented the deepest CMDs of Draco II to date. In Fig.~5a they show the CMD of Draco II up to the main sequence cutoff along with fits using theoretical isochrones. This analysis shows a handful of red-straggler stars to the red of the red giant branch, whose precise origin is presently unknown. These could be explained by Hawking stars in the maximal scheme as they end their lives in an extended, alternate post main sequence in which they occupy the red straggler space of CMDs. \mb{Hawking stars in the other two schemes presented, however, do not lead to red stragglers due to the rapid consumption of the star at the onset of canonical Bondi accretion, so it should be noted that this possibility is heavily accretion dependent.} In this section, we create synthetic stellar populations of a Draco II-like UFD galaxy including \mb{maximal scheme} Hawking stars in order to determine if they can explain the handful of red stragglers observed in Draco II today. 

\subsection{Initial Mass Function}\label{sec:IMF}

We begin by calculating the initial mass function for UFD galaxies. For these galaxies we assume the stellar population is formed in a single burst in the early universe. This also implies a low average metallicity which we take to be [Fe/H]~$=-2$ in this work when calculating both typical and Hawking star lifetimes, though this could be as low as [Fe/H]~$\approx -2.7$ \citep{draco_Z2016, draco_Z2018}. 

In this work, we adopt the widely used broken-power law distribution using values from \cite{Kroupa_2001}. The general form of this distribution is given by

\begin{equation}
    \dfrac{\textrm{d}N(\alpha_1,\alpha_2,M)}{\textrm{d}M}\propto
    \begin{cases}
        M^{\alpha_1} \text{ for }0.2~\textrm{M}_\odot\le M<0.5~\textrm{M}_\odot\\
        kM^{\alpha_2} \text{ for }M\ge0.5~\textrm{M}_\odot,
    \end{cases}
\label{eq:powerlaw}
\end{equation}

\noindent where $k=(0.5$\Msun)$^{\alpha_1 - \alpha_2}$ is a continuity constant and both $\alpha$ values are the power-law exponents. We use $\alpha_1 = -1.3$ and $\alpha_1 = -2.3 $ for the Kroupa IMF.\footnote{The IMF is known to be top heavy at lower metallicity; \cite{marks2012evidence} suggests $\alpha_1 = -0.3$ and $\alpha_1 = -1.3 $ for $\rm{[Fe/H]} =-2$. We also simulated with these values and the qualitative physics of our results below does not change, though the number of stars forming at 0.3~\Msun\ was noticeably reduced. }

We generate a stellar population by stochastically sampling the Kroupa IMF for an arbitrary number of stars in bins of width 0.1~\Msun. We sample from 0.3~\Msun\ to 100~\Msun, with the low end chosen due to numerical errors in calculating the lifetimes of Hawking stars with masses lower than 0.3~\Msun. The high end is arbitrary beyond roughly 6~\Msun\ as low star count galaxies like Draco II have low likelihood of forming higher mass stars, though this likelihood is still higher than that of the Milky Way for low metallicity galaxies. For convenience, we use the Python package Chempy. 

\subsection{Capture Probability}\label{sec:capture}

PBH capture occurs during the collapse of protostellar gas clouds after becoming gravitationally bound and sinking to the center due to dynamical friction \citep{esser2023impact}, so we take it into account only in the initial formation burst discussed above. This capture is an independent, stochastic process and therefore follows a Poisson distribution. The probability of a star capturing a PBH upon formation is then given by 

\begin{equation}
    P(N \geq 1) = 1 - e^{f \eta \nu(M)}
\end{equation}

\noindent where $f=1$ in this work and is the fraction of dark matter as PBHs, $\eta$ is the ``merit factor'' introduced by \cite{esser2023constraints}, and $\nu(M)$ is the mean number of captured PBHs as a function of stellar mass $M$ when $f=\eta=1$. For $\nu(M)$ we use the analytic fit

\begin{equation}
    \nu(M) = 3.76 \left ( \frac{M}{\textrm{M}_\odot} \right )^{0.685} - 0.883.
\end{equation}

\noindent The merit factor contains information about the dark matter properties of the host galaxy, and is given by

\begin{equation}
    \eta = \frac{\rho_{\rm{DM}}}{100~\rm{GeV/cm^3}} \left ( \frac{7~\rm{km~s^{-1}}}{\sqrt{2}\sigma_\nu} \right )^3
\end{equation}

\noindent where $\rho_{\rm{DM}}$ is the dark matter density and $\sigma_\nu$ is its velocity dispersion. For Draco II, \cite{draco_Z2018} shows a velocity dispersion at the 95\% confidence level of $<5.9~\rm{km/s}$, though this may be as low as $2.9~\rm{km/s}$. The velocity dispersion implies a half-light mass of about $\rm{log}_{10}(M_{1/2})=5.5$ in solar units, which we use along with the half-light radius $r_{1/2}=19$~pc to obtain the DM density $\rho_{\rm{DM}}=418~\rm{GeV/cm^{3}}$, assuming that baryonic matter is subdominant \citep{draco_Z2016, esser2023constraints, esser2023impact}.

\subsection{Monte Carlo Simulation}

\begin{figure*}[t]\label{fig:hr}
    \centering
    \includegraphics[trim={0 5cm 0 3.5cm},clip,width=0.95\textwidth]{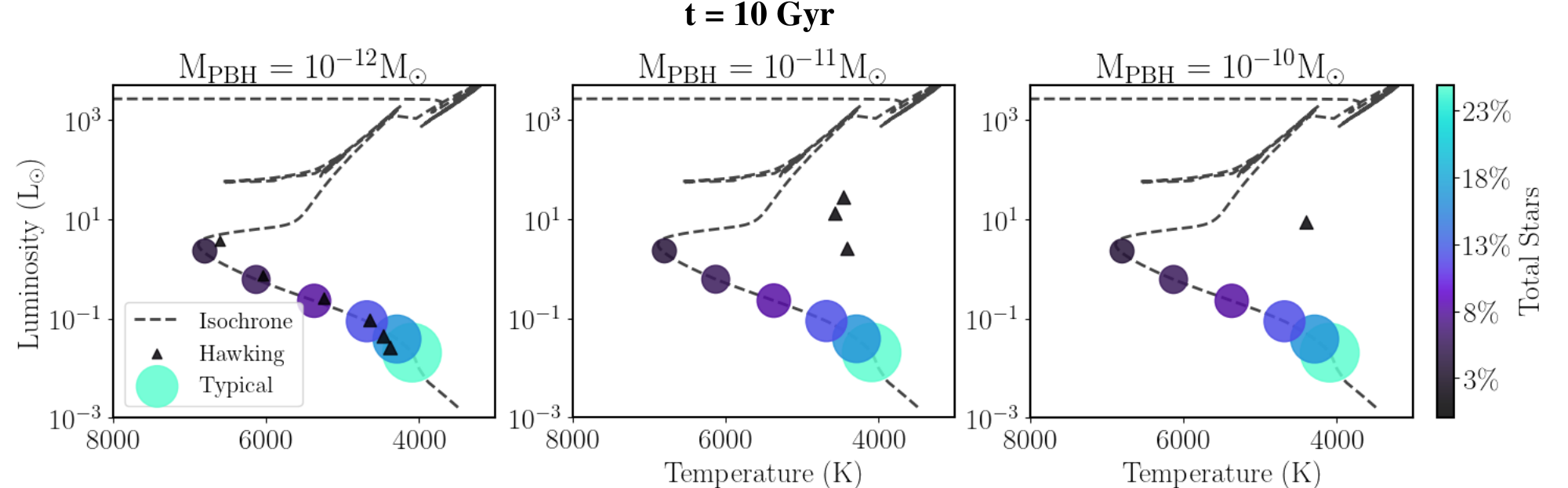}
     \vspace{2.5 mm} \hrule \vspace{2.5 mm} 
    \includegraphics[trim={0 5cm 0 3.5cm},clip,width=0.95\textwidth]{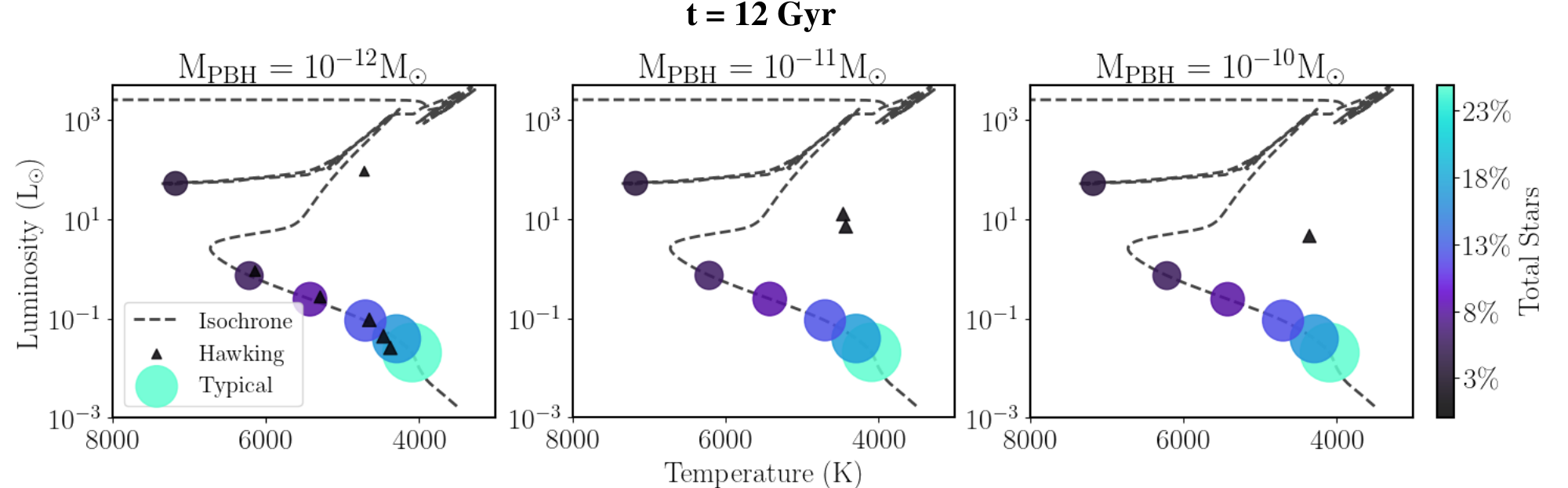}
     \vspace{2.5 mm} \hrule \vspace{2.5 mm} 
    \includegraphics[trim={0 5cm 0 3.5cm},clip,width=0.95\textwidth]{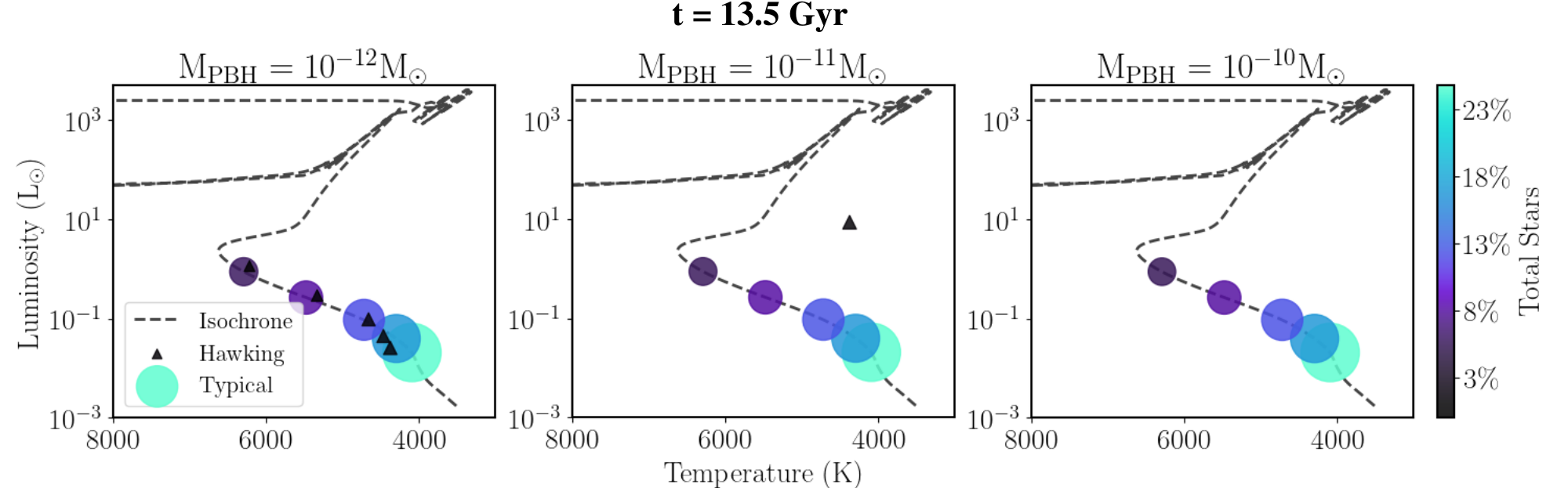}
    \caption{HR diagrams of Draco II-like population syntheses at 10~Gyr(top), 12~Gyr (middle), and the present age of Draco II 13.5~Gyr (bottom). We include monochromatic PBH mass distributions of $10^{-12}$ (left), $10^{-11}$ (middle), and $10^{-10}$~\Msun\ (right). Typical stars without a central PBH are indicated by circles and Hawking stars by triangles. The size and color scale with the percentage of total stars that the point makes up. An isochrone fit from \textsc{Mist} is shown by the dashed line. We include animations showing the evolution in time in the supplemental materials.}
    \label{fig:draco_hr}
\end{figure*}

We now generate a synthetic UFD galaxy. To do so, we stochastically sample the \cite{Kroupa_2001} IMF as discussed in Sec.~\ref{sec:IMF} to get the total stellar population. The number of stars generated is arbitrary as the number of captures scales with star count, so results will be presented as a percentage of total population. We then designate a fraction of the population to be Hawking stars using the probabilities discussed in Sec.~\ref{sec:capture} and keep counts for each in 0.1~\Msun\ bins. This counting process is repeated for 100 trials, and the counts for each mass bin are then averaged to give the final counts in each. We then get surface temperatures and luminosities for each Hawking star from the numerical calculations discussed in Sec.~\ref{sec:SEM} at times up to 13.67 Gyr in 0.33 Gyr increments. We do the same for the typical stars using data from the \textsc{Mist} stellar tracks \citep{MIST_1, MIST_2}. We exclude any stars that have been either transmuted to black holes or become white dwarfs by the respective time. We repeat this process for $10^{-12}$, $10^{-11}$, and $10^{-10}$~\Msun\ PBH seed masses, assuming a monochromatic PBH mass distribution. \cite{draco_Z2018} reports the age of Draco II to be about 13.5~Gyr, so we include HR plots at that time for each PBH mass in Fig.~\ref{fig:hr}. The age uncertainty may be about a Gyr, so we also include HR diagrams at 10 Gyr and 12 Gyr, with animations showing the full time evolution of the galaxy in the supplemental materials.

At 13.5 Gyr in age, shown in Fig.~\ref{fig:draco_hr}, a Draco II-like galaxy with PBHs of $\rm{M_{PBH} = 10^{-12}}$~\Msun\ results in very few, if any, red stragglers as shown by the left-most plot of Fig.~\ref{fig:hr}. All Hawking stars involved with this PBH mass ascend the giant branch as typical stars and only transmute to black holes once they have become white dwarfs. However, it should be noted that their path along the giant branch is slightly to the red but within error and not as much as the Hawking stars with a novel post main sequence. Any lower PBH masses would be expected to produce similar evolution (\textit{i.e.} these stars are all in the `red' region of the lifetime analysis presented in Fig.~\ref{fig:bondipepsi}). 

Increasing the PBH mass to $10^{-11}$~\Msun\ we see a significant increase in the number of red stragglers beginning at about 7~Gyr. This overproduction is later corrected when the majority of red stragglers are transmuted to black holes by the current age of the universe, resulting in the correct order of the sparse red straggler count found in observation \citep{weisz2023jwst}. 
This also implies a peak production mass somewhere between $10^{-12}$ and $10^{-11}$~\Msun\ where red stragglers are overly abundant by the current age of the UFD. If the PBH mass is $10^{-10}$~\Msun, the upper limit of the asteroid mass window, we see a similar but even earlier overproduction time resulting in all Hawking stars being transmuted to black holes well before the present (i.e., these stars would all be found in the `blue' region of the lifetimes in Fig.~\ref{fig:bondipepsi}). Therefore, Hawking stars may be able to provide the observed excess of red stragglers in UFDs if PBHs near $10^{-11}$ \Msun\ are a large fraction of the dark matter. 

\section{Conclusions}

In this work we have analytically calculated the lifetimes of sub-Chandrasekhar mass Hawking stars as a percentage of their main sequence lifetimes without a central black hole using three different accretion schemes. We have also verified these estimates with numerical simulations over a range of stellar and PBH masses and found our analytic lifetimes agree with stellar evolution models to less than a factor of two, with the error dominated by stellar metallicity. Additionally we have extended these numerical simulations to lower metallicities that can now be used for more detailed population syntheses. 

Despite uncertainty in the accretion physics, many stars above about 0.8~\Msun\ can be consumed within the age of the Universe regardless of PBH mass. For some, this is because core contraction at the end of the canonical main sequence increases the core density and conductivity, accelerating the accretion rate and cutting short their time in the red giant branch. For others, it is simply from the rapid consumption of the high density white dwarf formed in the typical post main sequence. The factor of two sensitivity of the lifetime to metallicity is especially important as star forming halos in the early universe should have lower masses and thus higher capture rates. Even in our maximal efficiency scheme, with the lowest accretion rate, early low mass Hawking stars could be converted into stellar mass black holes within a Hubble time. In the minimal efficiency scheme, nearly all sub-Chandrasekhar mass stars would have their main sequence lifetimes cut short by asteroid-mass PBHs. More generally, these PBHs would have transmuted all stellar masses to BHs within the age of the Universe. It suffices to say that if asteroid mass PBHs both exist and comprise a significant portion of dark matter, Hawking stars would be a viable formation pathway for sub-Chandrasekhar mass black holes. 

Motivated by the observation of a handful of red straggler stars in the Draco II UFD galaxy, we have generated a synthetic Draco II-like population containing Hawking stars with a range of PBH seed masses accreting in our maximal efficiency scheme. We show that there is a PBH mass between $10^{-12}$~\Msun\ and $10^{-11}$~\Msun\ where red stragglers are overproduced. Below this peak, none of them evolve to red stragglers early and transmute rapidly to stellar mass BHs at the end of their typical main sequence lifetimes. Near this point, our model is capable of producing an appropriate number of red stragglers at the current age of Draco II. However, for PBH masses an order of magnitude or more above this limit, close to the upper limit of the asteroid mass window, the red stragglers are overproduced early and convert to stellar mass BHs well before the current age of the universe.

We emphasize that large uncertainties remain in the accretion physics and the modeling of Draco II. Capture rates depend on the dark matter density and velocity dispersion at early times, while we only have measurements of a highly evolved population today. Likewise, extended density and velocity profiles could be used to improve modeling as well. 
Theoretical improvements to capture rates and accretion are also obviously needed before making detailed population synthesis calculations to compare with observation.
We emphasize again that work calculating capture rates and black hole accretion physics is ongoing. Our work is agnostic and largely surveys a range of possible impacts of PBH dark matter on stellar populations, and demonstrates that the qualitative stellar populations of at least one UFD can be matched by PBH dark matter. 

\begin{acknowledgments}
The authors thank Nicholas Esser and Florian K\"{u}hnel for helpful discussions. 
Financial support for this publication comes from Cottrell Scholar Award CS-CSA-2023-139 sponsored by Research Corporation for Science Advancement. This work was supported by a grant from the Simons Foundation (MP-SCMPS-00001470) to MC. This research was supported in part by the National Science Foundation under Grant No. NSF PHY-1748958. 
M.C. thanks the KITP for hospitality and acknowledges support as a KITP Scholar. 

\end{acknowledgments}

\appendix

\begin{figure*}[ht]
    \label{fig:misc}
    \includegraphics[width=\textwidth]{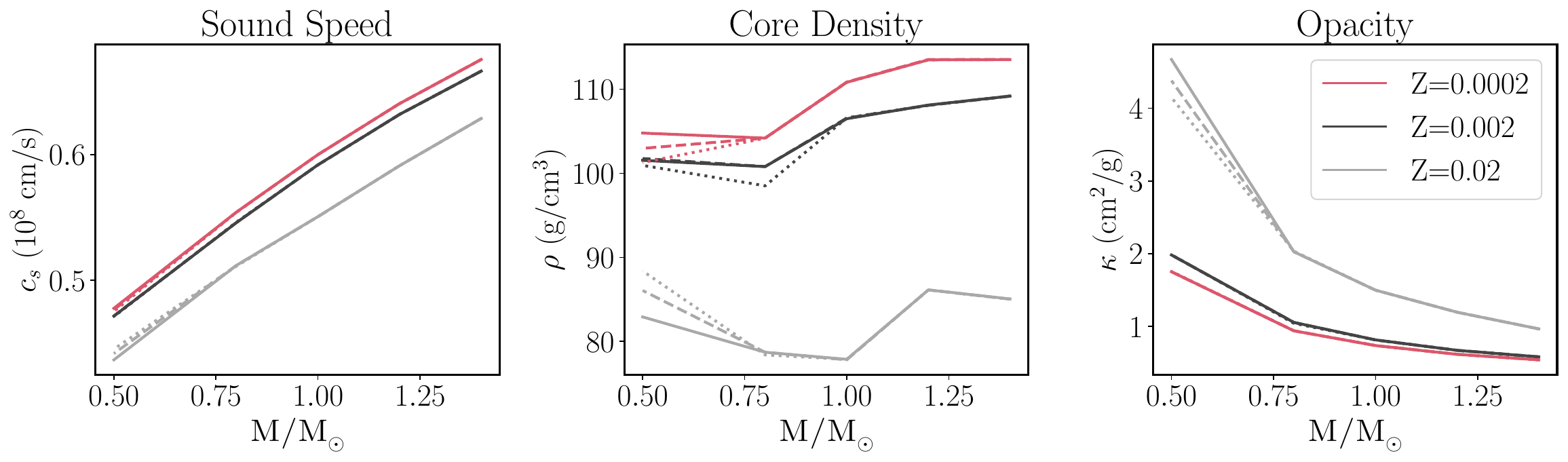}
    \caption{The ZAMS sound speed (left), density (middle), and opacity (right) for several Hawking star models. Models included are permutations of 100\%, 10\%, and 1\% solar metallicity and PBH seed masses $10^{-12}$~\Msun (solid), $10^{-11}$~\Msun (dashed), and $10^{-10}$~\Msun (dotted).}
\end{figure*}

In Fig.~\ref{fig:misc} we show the ZAMS values for core sound speed $c_s$, density $\rho$, and opacity $\kappa$ as a function of stellar mass for a number of Hawking star models. These apply to both the maximal and intermediate schemes discussed in Sec.~\ref{sec:acc}. However, these do not apply to the minimal models as many of them do not make it to ZAMS due to the rapid accretion rate. The strong dependence of $\kappa$ on stellar mass and metallicity is largely responsible for the lifetimes seen in the maximal model in Fig.~\ref{fig:num-lt}, while the variation in sound speed with stellar mass determines the lifespan of the minimal model. 




\end{document}